\documentclass[aps,superscriptaddress,nofootinbib,11pt]{revtex4-1} 
\pdfoutput=1

\usepackage[utf8]{inputenc}
\usepackage{amsmath}
\usepackage{amssymb}
\usepackage{graphicx}
\usepackage{dsfont}
\usepackage{multirow}
\usepackage[colorlinks=false,pdfborder={0 0 0}]{hyperref}

\allowdisplaybreaks

\def\MGUT{M_\mathrm{GUT}}
\def\SU{\mathrm{SU}}
\def\U{\mathrm{U}}
\def\e{\mathrm{e}}
\def\i{\mathrm{i}}
\def\Z{\mathbb{Z}}
\newcommand{\rep}[1]{\mathbf{#1}}
\newcommand{\repb}[1]{\mathbf{\overline{#1}}}

\begin{document}

\title{Flavor Symmetries in an SU(5) Model of Grand Unification}

\author{Malte Lindestam}
\email{malteli@kth.se}
\affiliation{Department of Physics,
	School of Engineering Sciences,
	KTH Royal Institute of Technology,
	AlbaNova University Center,
	Roslagstullsbacken 21,
	SE--106 91 Stockholm,
	Sweden}
	
\author{Tommy Ohlsson}
\email{tohlsson@kth.se}
\affiliation{Department of Physics,
	School of Engineering Sciences,
	KTH Royal Institute of Technology,
	AlbaNova University Center,
	Roslagstullsbacken 21,
	SE--106 91 Stockholm,
	Sweden}
\affiliation{The Oskar Klein Centre for Cosmoparticle Physics,
	AlbaNova University Center,
	Roslagstullsbacken 21,
	SE--106 91 Stockholm,
	Sweden}

\author{Marcus Pernow}
\email{pernow@kth.se}
\affiliation{Department of Physics,
	School of Engineering Sciences,
	KTH Royal Institute of Technology,
	AlbaNova University Center,
	Roslagstullsbacken 21,
	SE--106 91 Stockholm,
	Sweden}
\affiliation{The Oskar Klein Centre for Cosmoparticle Physics,
	AlbaNova University Center,
	Roslagstullsbacken 21,
	SE--106 91 Stockholm,
	Sweden}

\begin{abstract}
We investigate the options for imposing flavor symmetries on a minimal renormalizable non-supersymmetric $\SU(5)$ grand unified theory, without introducing additional flavor-related fields. Such symmetries reduce the number of free parameters in the model and therefore lead to more predictive models. We consider the Yukawa sector of the Lagrangian, and search for all possible flavor symmetries. As a result, we find 25 distinct realistic flavor symmetry cases, with $\Z_2$, $\Z_3$, $\Z_4$, and $\U(1)$ symmetries, and no non-Abelian cases.
\end{abstract}

\maketitle

\section{Introduction}\label{ch:intro}
Grand unified theories (GUTs) aim to embed the Standard Model (SM) gauge group $\SU(3)_C\times \SU(2)_L\times \U(1)_Y$ into a larger group at some high energy, called the GUT scale $\MGUT$. As a consequence, different SM fields are placed in the same representations. Thus, the different forces and fields of the SM are unified above $\MGUT$.
There are several candidate groups that GUTs may be based on, $\SU(5)$ being an appealing choice due to its simplicity. It is the smallest simple gauge group that contains the SM gauge group, and the SM fields fit neatly into $\SU(5)$ representations without needing to introduce additional fields.

However, the original $\SU(5)$ Georgi--Glashow model \cite{Georgi:1974sy} has several issues. It fails to actually unify the SM gauge couplings \cite{Marciano:1983gf} and it predicts fermion masses in conflict with experimentally measured values. Furthermore, it fails to provide an explanation for neutrino masses, like the SM. Therefore, extensions of the original model are required to construct a viable model. The different options for extending GUTs have been much studied in the literature. In $\SU(5)$ models, a $\mathbf{45}$ representation of scalars may be introduced to solve the problems of fermion masses and unification \cite{Georgi:1979df}.
By introducing new fields, various mechanisms for generating neutrino masses may be introduced to $\SU(5)$ models. These include the popular type I \cite{Minkowski:1977sc, Gell-Mann:1979vob, Mohapatra:1979ia, Sawada:1979dis, Schechter:1980gr} and type II \cite{Magg:1980ut, Mohapatra:1980yp} seesaw mechanisms.

While these amended models can generate the masses and mixings present in the SM, they generally do not provide any explanation for patterns in them, such as the observed fermion mass hierarchy. This is known as the flavor problem, and is present in the SM as well. In an attempt to solve this problem, one can impose additional symmetries on the fields of the model. By restricting the parameters of the model and imposing texture zeros on the Yukawa matrices, these so-called flavor symmetries can help provide natural explanations for why the experimentally observed masses and mixing parameters of the fermions obtain the values that they do. In refs. \cite{Ivanov:2015xss, Ferreira:2015jpa}, the possibilities for flavor symmetries in GUTs based on the group $\mathrm{SO}(10)$ have been studied. We have adapted and applied the methods developed in these works to a model based on the group $\SU(5)$ in order to explore the possible flavor symmetries that may be imposed on it.

Previous approaches to flavor in $\SU(5)$ models by imposing discrete symmetries may be found, for example, in Refs.~\cite{Chen:2007afa,Emmanuel-Costa:2013gia,Campos:2014lla,CarcamoHernandez:2014ety,Arbelaez:2015toa,Perez:2020nqq}. There are also other approaches to flavor, which generally introduce new fields. These include the Froggatt--Nielsen \cite{Froggatt:1978nt} and clockwork \cite{Alonso:2018bcg, Patel:2017pct,Sannino:2019sch} mechanisms.

This work is structured as follows. In Sec.~\ref{ch:model_specifics}, we outline the field content and Lagrangian of our model. Then, in Sec.~\ref{ch:flavor_symmetry}, we describe the flavor transformations and the method for deriving the possible symmetries. The results are given in Sec.~\ref{ch:cases}. In Sec.~\ref{ch:fits}, numerical fits are briefly discussed.  Finally, in Sec.~\ref{ch:conclusions}, we summarize and  conclude. In appendix~\ref{app:cases}, all the found flavor symmetry cases are listed. In appendix~\ref{app:NA}, the exclusion of non-Abelian cases is motivated.

\section{Lagrangian of the model}
\label{ch:model_specifics}

We study a non-supersymmetric renormalizable model with an additional scalar field in the $\mathbf{45}$ representation to solve the problems of down-type quark masses and gauge coupling unification \cite{Georgi:1979df}. To generate neutrino masses, a type II seesaw mechanism is introduced using a scalar in the $\mathbf{15}$ representation. The type II seesaw is suitable for the methods used in this study, more so than other mechanisms (such as the type I seesaw). The reason is that the type II seesaw only introduces a single new coupling matrix, which also has the feature of being directly proportional to the neutrino mass matrix. This simplifies calculations compared to other mechanisms. The type II mechanism is also appealing in that no additional fermion fields need to be introduced.

The fermions of the model are in the representations $\repb{5}_F^i$ and $\rep{10}_F^i$, where $i\in\{1,2,3\}$ labels the generation. In the Yukawa sector, the relevant scalars are $\rep{5}_S$, $\rep{45}_S$, and $\rep{15}_S$, with the first two generating charged fermion masses and the latter generating neutrino masses through the type II seesaw mechanism. For spontaneous symmetry breaking to the SM gauge group, we use a scalar in the adjoint representation $\rep{24}_S$.

The relevant part of the Lagrangian for our study is the Yukawa sector. It is given by \cite{Dorsner:2007fy}
\begin{align}
\label{yukawa_sector}
\begin{split}
    \mathcal{L}_Y = &- Y_1^{ij}  (\textbf{10}_F^i)^T_{\alpha\beta} C (\textbf{10}_F^j)_{\gamma\delta} (\textbf{5}_S)_{\varepsilon}\epsilon^{\alpha\beta\gamma\delta\varepsilon}
    - X_1^{ij} (\textbf{10}_F^i)^T_{\alpha\beta} C (\overline{\textbf{5}}_F^j)^\alpha (\textbf{5}_S)^{*\:\beta} \\
    &- Y_2^{ij} (\textbf{10}_F^i)^T_{\alpha\beta} C (\textbf{10}_F^j)_{\zeta\gamma} (\textbf{45}_S)_{\delta\varepsilon}^\zeta \epsilon^{\alpha\beta\gamma\delta\varepsilon}
    - X_2^{ij} (\textbf{10}_F^i)^T_{\alpha\beta} C (\overline{\textbf{5}}_F^j)^\gamma (\textbf{45}_S)^{*\:\alpha\beta}_{\gamma} \\
    &- N^{ij} (\mathbf{\overline{5}}_F^i)^T_\alpha C (\mathbf{\overline{5}}_F^j)_\beta (\mathbf{15}_S)^{\alpha\beta} + \mathrm{h.c.},
\end{split}
\raisetag{1\baselineskip}
\end{align}
where $\epsilon$ is the 5-index totally antisymmetric symbol, $C$ is the charge conjugation matrix, and $Y_{1,2}$, $X_{1,2}$ and $N$ are $3\times3$ coupling matrices in generation space. Generation indices are denoted by Latin letters, whereas Greek letters are used for $\SU(5)$ indices. In the terms where \textbf{10}$_F$ appears twice, one may rename indices such that the two are exchanged. One then finds that $Y_1$ must be symmetric and $Y_2$ antisymmetric \cite{Davidson:1980sx}. Doing the same with \textbf{5}$_F$, one finds that $N$ must be symmetric. No such constraint exists for $X_{1,2}$, since they couple to two different fermion multiplets.

The various scalars are not of much interest to this study, beyond their interactions with fermions. Thus, the scalar potential is not considered in detail. Its form is rather complicated, owing to the large number of scalars involved \cite{Eckert:1983bn}. Most of the scalars obtain large masses from their interactions with the $\mathbf{24}_S$, in general at the GUT scale. However, a single linear combination of the two scalar $\SU(2)_L$ doublets is assumed to survive at low energies to act as the SM Higgs doublet. This means that fine-tuning of the masses is necessary, as an extended version of the doublet-triplet splitting problem in the Georgi--Glashow model.

While the scalars in the theory obtain masses as $\SU(5)$ is broken (excluding the unphysical Goldstone modes absorbed by the massive gauge bosons), the fermions remain massless. When $\SU(3)_C\times \SU(2)_L\times \U(1)_Y$ is broken, they obtain masses through the usual SM Higgs mechanism. To break electroweak symmetry, the $\mathbf{5}_S$ obtains a vacuum expectation value (VEV) of
\begin{equation}
    \langle (\mathbf{5}_S)^5 \rangle = \frac{v_{5}}{\sqrt{2}}
\end{equation}
and the $\mathbf{45}_S$ obtains a VEV of
\begin{equation}
    \langle (\mathbf{45}_S)^{\alpha5}_\alpha \rangle = \frac{v_{45}}{\sqrt{2}} \:\:\: \mathrm{for} \:\:\: \alpha=1,2,3,
\end{equation}
\begin{equation}
    \langle (\mathbf{45}_S)^{45}_4 \rangle = -3\frac{v_{45}}{\sqrt{2}}.
\end{equation}
The $\mathbf{15}_S$ also obtains a VEV
\begin{equation}
    \langle (\mathbf{15}_S)^{55} \rangle = v_{15},
\end{equation}
which generates neutrino masses. All indices not mentioned have a VEV of zero. With these VEVs, the Yukawa sector results in mass terms for all fermions. After some computations, the mass matrices are given by \cite{Dorsner:2007fy}
\begin{align}
    M_u &= 4(Y_1 + Y_1^T)\frac{v_5}{\sqrt{2}} - 8(Y_2 - Y_2^T)\frac{v_{45}}{\sqrt{2}},\\
    M_d &= X_1\frac{v_5^*}{\sqrt{2}} + 2X_2\frac{v_{45}^*}{\sqrt{2}},\\
    M_e &= X_1^T\frac{v_5^*}{\sqrt{2}} - 6X_2^T\frac{v_{45}^*}{\sqrt{2}},\\
    M_\nu &= Nv_{15}.
\end{align}

Knowing experimental facts about the masses and mixings of the SM fermions, two requirements may be imposed on these coupling matrices. Firstly, all quarks and charged leptons are known to be massive. Thus, their mass matrices may not be singular. In contrast, the neutrino mass matrix may be singular, since one neutrino may be massless. Secondly, it is an experimental fact that there is mixing of all quark generations and all lepton generations. Performing the matrix multiplication, one finds that a generation is decoupled if all nonzero elements in a given row in $M_u$ and $M_d$ (or $M_e$ and $M_\nu$) are the only nonzero element in their respective columns. For example, the first quark generation is decoupled if the mass matrices take the form
\begin{equation}
M_u = 
    \begin{bmatrix}
0 & 0 & \times \\
\times & \times & 0 \\
\times & \times & 0
    \end{bmatrix}
    \:\:\:\:\:\:
    \mathrm{and}
    \:\:\:\:\:\:
M_d =
    \begin{bmatrix}
\times & 0 & \times \\
0 & \times & 0 \\
0 & \times & 0
    \end{bmatrix},
\end{equation}
where $\times$ represents nonzero values. Crucially, both of these requirements are independent of the renormalization group (RG) running of the mass matrices (the relevant beta functions may be found in ref.~\cite{Schmidt:2007nq}). That is to say, if they hold at some energy, they hold at all energies. This is important, since the model and its coupling matrices are defined at the GUT scale, while the experimental facts that motivate these requirements are only known at lower energies.

Finally, the assumption is made that the different generations of each kind of fermion have different masses at the GUT scale. While this is known to be true at lower energies, it may cease to be the case due to the RG running of the masses. However, RG running of fermion masses to the GUT scale using the SM results in all different masses \cite{Xing:2007fb, Huang:2020hdv}, which lends credence to this assumption.

\section{Flavor symmetry transformations}
\label{ch:flavor_symmetry}
The topic of this study is the possible flavor symmetries of the chosen model. We consider global symmetries under which the fields in the model transform as
\begin{align}
    \mathbf{\overline{5}}_F^i &\rightarrow W_5^{ij} \mathbf{\overline{5}}_F^j,
    \label{trans1}\\
    \mathbf{10}_F^i &\rightarrow W_{10}^{ij} \mathbf{10}_F^j,
    \label{trans2}\\
    \mathbf{5}_S &\rightarrow \e^{\i\varphi_5} \mathbf{5}_S,
    \label{trans3}\\
    \mathbf{45}_S &\rightarrow \e^{\i\varphi_{45}} \mathbf{45}_S,
    \label{trans4}\\
    \mathbf{15}_S &\rightarrow \e^{\i\varphi_{15}} \mathbf{15}_S,
    \label{trans5}
\end{align}
where $i,j\in\{1,2,3\}$ are generation indices, $W_5$ and $W_{10}$ are unitary $3\times3$ matrices, and $\varphi_{5}$, $\varphi_{45}$, and $\varphi_{15}$ are phases. The transformation may be discrete or continuous. A symmetry may have multiple sets of these transformation parameters, meaning that the symmetry is either a product of symmetries or non-Abelian.

If a model has a flavor symmetry, then the entire Lagrangian must be invariant under this symmetry. The Yukawa sector in eq.~\eqref{yukawa_sector} is of interest, since it is closely related to the fermion mass matrices, about which experimental facts are known. The terms in the Yukawa sector are invariant as long as
\begin{align}
    \e^{\i\varphi_5} W_{10}^T Y_1 W_{10} &= Y_1,
    \label{invariantY1}\\
    \e^{\i\varphi_{45}} W_{10}^T Y_2 W_{10} &= Y_2,
    \label{invariantY2}\\
    \e^{-\i\varphi_5} W_{10}^T X_1 W_5 &= X_1,
    \label{invariantX1}\\
    \e^{-\i\varphi_{45}} W_{10}^T X_2 W_5 &= X_2,
    \label{invariantX2}\\
    \e^{\i\varphi_{15}} W_5^T N W_5 &= N.
    \label{invariantN}
\end{align}
It is assumed that all $W$ matrices are simultaneously diagonalizable. This disallows non-Abelian symmetries, which is motivated in appendix~\ref{app:NA}. Since the generation bases for both the 5- and 10-plet are arbitrary, the bases where all $W$ matrices are diagonal are chosen.  As the $W$ matrices are unitary, they may then be expressed in terms of three phases each, i.e.
\begin{align}
    W_{5} &= \mathrm{diag}(\e^{\i\varphi_{5_F}^1}, \e^{\i\varphi_{5_F}^2}, \e^{\i\varphi_{5_F}^3}),
    \label{w5}\\
    W_{10} &= \mathrm{diag}(\e^{\i\varphi_{10_F}^1}, \e^{\i\varphi_{10_F}^2}, \e^{\i\varphi_{10_F}^3}).
    \label{w10}
\end{align}
This basis choice simplifies the situation greatly at no loss of generality.

Equations~\eqref{invariantY1}--\eqref{invariantN} impose restrictions on the phases $\varphi_5$, $\varphi_{10}$, $\varphi_{15}$, $\varphi_{45}$, $\varphi_{5_F}^i$, and $\varphi_{10_F}^i$. Exactly what these restrictions are depends on the forms of the coupling matrices. To investigate the different possible forms of the coupling matrices, we adapt and use method developed in refs.~\cite{Ivanov:2015xss, Ferreira:2015jpa}. For diagonal $W$ matrices, each matrix element in eqs.~\eqref{invariantY1}--\eqref{invariantN} may be treated separately. For invariance to hold, every element in every coupling matrix gives rise to an equation in the form
\begin{equation}
    \e^{\i\varphi}x = x,
\end{equation}
where $x$ is an element of a coupling matrix and $\varphi$ is a linear combination of the nine phases of the symmetry transformation. If the element $x$ is nonzero, then the phase $\varphi$ must be a multiple of $2\pi$. The equations corresponding to nonzero elements may all be collectively written in matrix form as
\begin{equation}
    \mathbf{D} \vec{\Phi} = 2\pi \vec{N},
\end{equation}
where $\vec{N}$ is a vector of arbitrary integers and $\vec{\Phi}$ is a vector of the nine phases of the symmetry: $\varphi_{5}$, $\varphi_{45}$, $\varphi_{15}$, and the six phases in eqs.~\eqref{w5} and \eqref{w10}. The equations for each nonzero element determine the matrix $\mathbf{D}$, which only has integer elements. The solutions to this system of equations are not interesting by themselves, as they merely correspond to possible transformations. The crucial information is the structure of these solutions, and the symmetries they imply. This becomes clearer if $\mathbf{D}$ is transformed into its Smith normal form (SNF) $\mathbf{D}_{\mathrm{SNF}} = \mathbf{L}\mathbf{D}\mathbf{R}$, where $\mathbf{L}$ and $\mathbf{R}$ are invertible matrices with integer elements. A matrix in SNF is diagonal with integer elements, where each diagonal element is a divisor of the next one. The system can be reformulated as
\begin{equation}
    \mathbf{D}_{\mathrm{SNF}} \Tilde{\vec{\Phi}} = 2\pi \Tilde{\vec{N}},
\end{equation}
where $\Tilde{\vec{\Phi}} = \mathbf{R}^{-1}\vec{\Phi}$ is a new vector of phases and $\Tilde{\vec{N}}=\mathbf{L}\vec{N}$ is another vector of arbitrary integers. Each phase in $\Tilde{\vec{\Phi}}$ multiplied by its corresponding diagonal element in $\mathbf{D}_{\mathrm{SNF}}$ must be an integer times $2\pi$. If the diagonal element is 1, then the only unique value the corresponding phase can assume is 0. If the diagonal element is $d>1$, then the corresponding phase may assume $d$ different equally spaced unique values that are all invariant, which is indicative of a $\Z_d$ symmetry. If the diagonal element is 0, then the phase can assume any value, which corresponds to a $\U(1)$ symmetry. In this way, the symmetry available to each phase may be read off from the diagonal elements of $\mathbf{D}_{\mathrm{SNF}}$. Once the allowed values of $\Tilde{\vec{\Phi}}$ are determined, the allowed values of the original transformation parameters are obtained using
\begin{equation}
    \vec{\Phi} = \mathbf{R}\Tilde{\vec{\Phi}}.
\end{equation}

As an example of this process, consider the possible symmetries of a matrix in the form
\begin{equation}
Y_1 =
\begin{bmatrix}
\times & 0 & 0 \\
0 & \times & \times \\
0 & \times & \times
\end{bmatrix}.
\end{equation}
For the sake of simplicity, only this coupling matrix is considered and only eq.~\eqref{invariantY1} is used. Transformations are then described by only four phases, which may be put in a vector $\Phi$. Carrying out the method using the SNF, the only allowed values of $\Phi$ are found to be
\begin{equation}
    \vec{\Phi} =
    \begin{bmatrix}
    \varphi_{10_F}^1\\
    \varphi_{10_F}^2\\
    \varphi_{10_F}^3\\
    \varphi_5
    \end{bmatrix}
    =
    \begin{bmatrix}
    z\pi + \varphi\\
    \varphi\\
    \varphi\\
    -2\varphi
    \end{bmatrix},
\end{equation}
where $z=0,1$ and $\varphi$ may assume any value. This is a $\Z_2 \times \U(1)$ symmetry. The transformation can be expressed as two sets of parameters, i.e.
\begin{align}
\label{example_symmetry_1}
    W_{10} = \mathrm{diag}(-1,1,1)&, \:\:\:\:\:\: \e^{\i\varphi_5} = 1,\\
\label{example_symmetry_2}
    W_{10} = \mathrm{diag}(\e^{\i\varphi}, \e^{\i\varphi}, \e^{\i\varphi})&, \:\:\:\:\:\: \e^{\i\varphi_5} = \e^{-2\i\varphi},
\end{align}
one set for each symmetry.

In the chosen model, it is always possible to impose a $\U(1)$ symmetry on the Yukawa sector. If all generations of the $\mathbf{10}_F^i$ have charge $1$, all generations of the $\mathbf{\overline{5}}_F^i$ have charge $-3$, the $\mathbf{5}_S^i$ and the $\mathbf{45}_S^i$ have charge $-2$, and the $\mathbf{15}_S^i$ has charge $6$, all terms in the Yukawa sector are invariant, no matter the shape of the coupling matrices. Since this symmetry is always available and does not restrict the coupling matrices in any way, it is considered trivial and will not be considered further. Thus, the symmetry of the example in eqs.~\eqref{example_symmetry_1} and \eqref{example_symmetry_2} would be described as merely $\Z_2$.

\section{Flavor symmetry cases}
\label{ch:cases}
Clearly, the coupling matrices affect the possibilities for flavor symmetry. More specifically, what matters is which elements of the coupling matrices are nonzero. All possible combinations of coupling matrices can be divided into cases depending on which elements are nonzero. The cases may have different symmetries imposed on them, which can be found using the methods in section~\ref{ch:flavor_symmetry}. Every case that has a possible nontrivial symmetry will be called a (flavor) symmetry case. 

Many symmetry cases are transparently related to each other, and should not be considered distinct from each other. Since the bases are arbitrary, permutations of the generations of both the 10- and 5-plet do not affect the symmetry. Cases related in such a way are therefore considered to be equivalent.

If a case is the same as another one except for some of its elements being set to zero, then it will be called a subcase of the other case. If a case has a symmetry, then all of its subcases have that symmetry as well. However, since the subcase has fewer nonzero elements, there are fewer restrictions on its transformations and it is possible that it has a larger symmetry. For instance, a $\Z_2$ case may have $\Z_2\times \Z_2$ or $\U(1)$ subcases. If the symmetry of a case is the same as that of its subcase, the subcase is not considered distinct. Subcases with larger symmetries are however considered distinct.

Since there exists a prohibitively large number of cases to be checked, several shortcuts have to be used. The $Y_1$ and $X_1$ matrices couple to the $\mathbf{5}_S$, while the $Y_2$ and $X_2$ matrices couple to the $\mathbf{45}_S$. As found in ref.~\cite{Ivanov:2015xss}, there are only two options for these two sets that can give rise to distinct symmetry cases. The first option is that they do not overlap, i.e.~they have no nonzero elements in the same positions. The second option is that the two sets have nonzero elements in the exact same positions (except that $Y_2$ does not have any diagonal elements present in $Y_1$). All other cases, where the sets partially overlap, will be non-distinct subcases of other cases.

Furthermore, consider that all possible distinct symmetry cases for a subset of the coupling matrices (only the $Y_{1,2}$ matrices, for instance) are known. Then, in all distinct symmetry cases for the full set, the subset must take the form of one of the previously found partial symmetry cases or be entirely nonzero.
This allows the different coupling matrices to be considered one after the other instead of in nested loops, which speeds up calculations greatly. In practice, the choice was made to analyze $Y_1$ and $Y_2$ first, followed by $X_1$ and $X_2$, and finally $N$.

Using the mentioned shortcuts, all combinations of coupling matrices that could produce distinct flavor symmetry cases are evaluated. Imposing the three requirements on the mass matrices mentioned in section~\ref{ch:model_specifics}, only 25 cases remain. These correspond to the symmetries $\Z_2$, $\Z_3$, $\Z_4$, and U(1) and may be found in appendix~\ref{app:cases}. Non-Abelian cases were found to be irreconcilable with the requirements, as detailed in appendix~\ref{app:NA}. As an example, the flavor symmetry case denoted $A_2$ has the following coupling matrices and transformation parameters:
\begin{equation}
Y_1 =
\begin{bmatrix}
\times & \times & \times \\
\times & \times & \times \\
\times & \times & \times
\end{bmatrix},
\:\:\:\:\:
Y_2 =
\begin{bmatrix}
0 & 0 & 0 \\
0 & 0 & 0 \\
0 & 0 & 0
\end{bmatrix},
\:\:\:\:\:
X_1 =
\begin{bmatrix}
\times & 0 & 0 \\
\times & 0 & 0 \\
\times & 0 & 0
\end{bmatrix},
\:\:\:\:\:
X_2 =
\begin{bmatrix}
0 & \times & \times \\
0 & \times & \times \\
0 & \times & \times
\end{bmatrix},
\:\:\:\:\:
N =
\begin{bmatrix}
\times & 0 & 0 \\
0 & \times & \times \\
0 & \times & \times
\end{bmatrix},
\end{equation}
\begin{equation}
    W_{10} = \mathrm{diag}(1,1,1), \:\:\:\:\:\: W_{5} = \mathrm{diag}(1,-1,-1),
\end{equation}
\begin{equation}
    \e^{\i\varphi_{5}} = 1, \:\:\:\:\:\: \e^{\i\varphi_{45}} = -1, \:\:\:\:\:\: \e^{\i\varphi_{15}} = 1.
\end{equation}

By relaxing the assumption made and allowing different particles to have the same mass at the GUT scale, additional cases are allowed. This leads to a total of 96 Abelian cases, again corresponding to the symmetries $\Z_2$, $\Z_3$, $\Z_4$, and U(1). By arguments similar to those in appendix~\ref{app:NA}, one can show that no non-Abelian cases are possible even without this assumption.

\section{Numerical Fits}
\label{ch:fits}

Numerically studying the feasibility of the 25 found flavor symmetry cases and fitting them to experimental data for masses and mixings is a computationally expensive task. Since the coupling matrices are defined at the GUT scale, but experimental data only known at lower energies, the RG running of the model must be performed. However, this is complicated by the fact that the scalar triplet (involved in the type II seesaw mechanism) has an intermediate-scale mass and must be integrated out at the energy scale corresponding to its mass. This affects the RG running of the lepton Yukawa matrices and gauge couplings, which in turn affects the running of all other Yukawa couplings \cite{Schmidt:2007nq}. This means that its mass scale must be included as a parameter in the fits. As such, one cannot perform the RG running prior to performing the fit, but must solve the RGEs for every sampled point in parameter space. While it can be carried out \cite{Ohlsson:2019sja}, doing so was judged to be beyond the scope of this work due to its large computational complexity.

To facilitate less computationally demanding fits, several assumptions were made. Mass data were taken from table~IV in ref.~\cite{Xing:2007fb}, where RG running has been performed to obtain the value of the masses at the GUT scale, while low-energy mixing data (from refs.~\cite{Charles:2004jd, ckmfitter_site, Esteban:2020cvm, nufit_site}) were assumed to hold at the GUT scale as well, as in, for example, refs.~\cite{Joshipura:2011nn,Altarelli:2013aqa,Dueck:2013gca}. In this manner, approximate RG running, neglecting the contribution of the scalar triplet, has been taken into account prior to carrying out the fits.

Of the 25 found cases, ten are not subcases of other cases. Since subcases will obtain fits of equal quality or worse as their parent cases, these ten cases were analyzed first. If they do not fit experimental data well, neither will their subcases. Preliminary fits were performed with uncertainties set to a minimum of 5 \%, wherein the cases denoted $D_1$, $D_2$, $D_3$, and $D_4$ were found to have decidedly poor fits, with $\chi^2$-values above 300. These cases were therefore discarded. Fits were then performed on the remaining six cases, with $\chi^2$-values shown in table~\ref{tab:fit}.

\begin{table}[!htb]
\begin{center}
\begin{tabular}{|c|c c|}
    \hline
    Case & $\chi^2$ (NO) & $\chi^2$ (IO) \\
    \hline
    $A_1$ & 69.02 & 73.31 \\
    $A_2$ & 7.981 & 7.870 \\
    $B_1$ & 83.89 & 83.89 \\
    $B_2$ & 37.62 & 37.62 \\
    $C_1$ & 87.72 & 87.65 \\
    $C_2$ & 82.71 & 82.76\\
    \hline
\end{tabular}
\end{center}
\caption{The $\chi^2$-values of the best fits found for the cases that are not subcases, excluding cases $D_1$, $D_2$, $D_3$, and $D_4$, which were found to be infeasible in preliminary fits. The abbreviation NO (IO) indicates normal (inverted) neutrino mass ordering.}
\label{tab:fit}
\end{table}

The only cases that can be said to fit the data well are the one denoted $A_2$, displayed in eqs.~\eqref{A2_matrices}--\eqref{A2_scalars}, as well as its subcase denoted $A_2'$, displayed in eqs.~\eqref{eq:A2pmats}--\eqref{eq:A2pscals}. The only difference between the two is a shift in the absolute neutrino mass scale, and therefore, this subcase can fit the data equally well as case $A_2$. Thus, of the 25 found cases, good fits were found only for case $A_2$ and case $A_2'$. An interesting feature of these cases is that they forbid the coupling between the $\mathbf{10}_F$ and $\mathbf{45}_S$. This coupling is a major contributor to proton decay in $\SU(5)$ models \cite{Langacker:1980js, Senjanovic:2009kr, Nath:2006ut}, and its exclusion can help raise the predicted proton lifetime~\cite{Dorsner:2006dj}.

\section{Summary and Conclusions}
\label{ch:conclusions}
As an adaptation of the work performed in refs. \cite{Ivanov:2015xss, Ferreira:2015jpa}, we have constructed a minimal $\SU(5)$ GUT upon which flavor symmetries may be imposed. This model includes no new flavor-related fields, instead exhausting the options for how the pre-existing fields may transform under new symmetries. In order to eliminate cases in conflict with experimental data, several restrictions have been set. Specifically, all charged fermions are required to have nonzero mass, mixing of all generations of quarks and leptons is imposed and it is assumed that the different generations of each kind of fermion all have different mass at the GUT scale. As a result, we have found that there are only 25 distinct options for flavor symmetry, listed in appendix~\ref{app:cases}. Notably, only Abelian symmetries have been found, namely $\Z_2$, $\Z_3$, $\Z_4$, and $\U(1)$ symmetries. Numerical fits using simplifying assumptions favor two of these options, denoted $A_2$ and $A_2'$ in appendix~\ref{app:cases}.

A possible future area of research is to numerically study these found flavor symmetry cases and how well they can fit experimental data, while taking into account the intermediate scale thresholds due to the scalar triplet. This would entail performing RG running down from the GUT scale in every iteration of the fitting procedure. Through such investigations, a more definitive verdict could be passed on the feasibility of the found cases.

\begin{acknowledgements}
T.O. acknowledges support by the Swedish Research Council (Vetenskapsrådet) through contract No. 2017-03934.
\end{acknowledgements}

\appendix

\section{Found symmetry cases}
\label{app:cases}

The 25 flavor symmetry cases found in section~\ref{ch:cases} are listed. Similar cases are named with the same letter, distinguished by subscripts. A superscript denotes that the case is a subcase of another case. The symmetries of the cases are characterized by the transformation parameters $W_{10}$, $W_5$, $\varphi_{5}$, $\varphi_{45}$ and $\varphi_{15}$ as defined in  eqs.~\eqref{trans1}--\eqref{trans5}. These are listed for each symmetry case, along with the shapes of the coupling matrices, where $\times$ represents a nonzero element. Cases with $\U(1)$ symmetry allow a continuous set of transformations, characterized by a phase $\phi$.\\

\noindent
Case $A_1$ ($\Z_2$ symmetry):
\begin{equation}
Y_1 =
\begin{bmatrix}
\times & \times & \times \\
\times & \times & \times \\
\times & \times & \times
\end{bmatrix},
\:\:\:\:\:
Y_2 =
\begin{bmatrix}
0 & 0 & 0 \\
0 & 0 & 0 \\
0 & 0 & 0
\end{bmatrix},
\:\:\:\:\:
X_1 =
\begin{bmatrix}
0 & \times & \times \\
0 & \times & \times \\
0 & \times & \times
\end{bmatrix},
\:\:\:\:\:
X_2 =
\begin{bmatrix}
\times & 0 & 0 \\
\times & 0 & 0 \\
\times & 0 & 0
\end{bmatrix},
\:\:\:\:\:
N =
\begin{bmatrix}
\times & 0 & 0 \\
0 & \times & \times \\
0 & \times & \times
\end{bmatrix},
\end{equation}
\begin{equation}
    W_{10} = \mathrm{diag}(1,1,1), \:\:\:\:\:\: W_{5} = \mathrm{diag}(-1,1,1),
\end{equation}
\begin{equation}
    \e^{\i\varphi_{5}} = 1, \:\:\:\:\:\: \e^{\i\varphi_{45}} = -1, \:\:\:\:\:\: \e^{\i\varphi_{15}} = 1.
\end{equation}

\noindent
Case $A_2$ ($\Z_2$ symmetry):
\begin{equation}
\label{A2_matrices}
Y_1 =
\begin{bmatrix}
\times & \times & \times \\
\times & \times & \times \\
\times & \times & \times
\end{bmatrix},
\:\:\:\:\:
Y_2 =
\begin{bmatrix}
0 & 0 & 0 \\
0 & 0 & 0 \\
0 & 0 & 0
\end{bmatrix},
\:\:\:\:\:
X_1 =
\begin{bmatrix}
\times & 0 & 0 \\
\times & 0 & 0 \\
\times & 0 & 0
\end{bmatrix},
\:\:\:\:\:
X_2 =
\begin{bmatrix}
0 & \times & \times \\
0 & \times & \times \\
0 & \times & \times
\end{bmatrix},
\:\:\:\:\:
N =
\begin{bmatrix}
\times & 0 & 0 \\
0 & \times & \times \\
0 & \times & \times
\end{bmatrix},
\end{equation}
\begin{equation}
    W_{10} = \mathrm{diag}(1,1,1), \:\:\:\:\:\: W_{5} = \mathrm{diag}(1,-1,-1),
\end{equation}
\begin{equation}
    \label{A2_scalars}
    \e^{\i\varphi_{5}} = 1, \:\:\:\:\:\: \e^{\i\varphi_{45}} = -1, \:\:\:\:\:\: \e^{\i\varphi_{15}} = 1.
\end{equation}

\noindent
Case $B_1$ ($\Z_2$ symmetry):
\begin{equation}
Y_1 =
\begin{bmatrix}
\times & 0 & 0 \\
0 & \times & \times \\
0 & \times & \times
\end{bmatrix},
\:\:\:\:\:
Y_2 =
\begin{bmatrix}
0 & \times & \times \\
\times & 0 & 0 \\
\times & 0 & 0
\end{bmatrix},
\:\:\:\:\:
X_1 =
\begin{bmatrix}
0 & 0 & 0 \\
\times & \times & \times \\
\times & \times & \times
\end{bmatrix},
\:\:\:\:\:
X_2 =
\begin{bmatrix}
\times & \times & \times \\
0 & 0 & 0 \\
0 & 0 & 0
\end{bmatrix},
\:\:\:\:\:
N =
\begin{bmatrix}
\times & \times & \times \\
\times & \times & \times \\
\times & \times & \times
\end{bmatrix},
\end{equation}
\begin{equation}
    W_{10} = \mathrm{diag}(-1,1,1), \:\:\:\:\:\: W_{5} = \mathrm{diag}(1,1,1),
\end{equation}
\begin{equation}
    \e^{\i\varphi_{5}} = 1, \:\:\:\:\:\: \e^{\i\varphi_{45}} = -1, \:\:\:\:\:\: \e^{\i\varphi_{15}} = 1.
\end{equation}

\noindent
Case $B_2$ ($\Z_2$ symmetry):
\begin{equation}
Y_1 =
\begin{bmatrix}
\times & 0 & 0 \\
0 & \times & \times \\
0 & \times & \times
\end{bmatrix},
\:\:\:\:\:
Y_2 =
\begin{bmatrix}
0 & \times & \times \\
\times & 0 & 0 \\
\times & 0 & 0
\end{bmatrix},
\:\:\:\:\:
X_1 =
\begin{bmatrix}
\times & \times & \times \\
0 & 0 & 0 \\
0 & 0 & 0
\end{bmatrix},
\:\:\:\:\:
X_2 =
\begin{bmatrix}
0 & 0 & 0 \\
\times & \times & \times \\
\times & \times & \times
\end{bmatrix},
\:\:\:\:\:
N =
\begin{bmatrix}
\times & \times & \times \\
\times & \times & \times \\
\times & \times & \times
\end{bmatrix},
\end{equation}
\begin{equation}
    W_{10} = \mathrm{diag}(-1,1,1), \:\:\:\:\:\: W_{5} = \mathrm{diag}(-1,-1,-1),
\end{equation}
\begin{equation}
    \e^{\i\varphi_{5}} = 1, \:\:\:\:\:\: \e^{\i\varphi_{45}} = -1, \:\:\:\:\:\: \e^{\i\varphi_{15}} = 1.
\end{equation}

\noindent
Case $C_1$ ($\Z_2$ symmetry):
\begin{equation}
Y_1 =
\begin{bmatrix}
\times & 0 & 0 \\
0 & \times & \times \\
0 & \times & \times
\end{bmatrix},
\:\:\:\:\:
Y_2 =
\begin{bmatrix}
0 & \times & \times \\
\times & 0 & 0 \\
\times & 0 & 0
\end{bmatrix},
\:\:\:\:\:
X_1 =
\begin{bmatrix}
0 & \times & \times \\
\times & 0 & 0 \\
\times & 0 & 0
\end{bmatrix},
\:\:\:\:\:
X_2 =
\begin{bmatrix}
\times & 0 & 0 \\
0 & \times & \times \\
0 & \times & \times
\end{bmatrix},
\:\:\:\:\:
N =
\begin{bmatrix}
\times & 0 & 0 \\
0 & \times & \times \\
0 & \times & \times
\end{bmatrix},
\end{equation}
\begin{equation}
    W_{10} = \mathrm{diag}(-1,1,1), \:\:\:\:\:\: W_{5} = \mathrm{diag}(1,-1,-1),
\end{equation}
\begin{equation}
    \e^{\i\varphi_{5}} = 1, \:\:\:\:\:\: \e^{\i\varphi_{45}} = -1, \:\:\:\:\:\: \e^{\i\varphi_{15}} = 1.
\end{equation}

\noindent
Case $C_2$ ($\Z_2$ symmetry):
\begin{equation}
Y_1 =
\begin{bmatrix}
\times & 0 & 0 \\
0 & \times & \times \\
0 & \times & \times
\end{bmatrix},
\:\:\:\:\:
Y_2 =
\begin{bmatrix}
0 & \times & \times \\
\times & 0 & 0 \\
\times & 0 & 0
\end{bmatrix},
\:\:\:\:\:
X_1 =
\begin{bmatrix}
\times & 0 & 0 \\
0 & \times & \times \\
0 & \times & \times
\end{bmatrix},
\:\:\:\:\:
X_2 =
\begin{bmatrix}
0 & \times & \times \\
\times & 0 & 0 \\
\times & 0 & 0
\end{bmatrix},
\:\:\:\:\:
N =
\begin{bmatrix}
\times & 0 & 0 \\
0 & \times & \times \\
0 & \times & \times
\end{bmatrix},
\end{equation}
\begin{equation}
    W_{10} = \mathrm{diag}(-1,1,1), \:\:\:\:\:\: W_{5} = \mathrm{diag}(-1,1,1),
\end{equation}
\begin{equation}
    \e^{\i\varphi_{5}} = 1, \:\:\:\:\:\: \e^{\i\varphi_{45}} = -1, \:\:\:\:\:\: \e^{\i\varphi_{15}} = 1.
\end{equation}

\noindent
Case $D_1$ ($\Z_3$ symmetry):
\begin{equation}
Y_1 =
\begin{bmatrix}
\times & 0 & 0 \\
0 & 0 & \times \\
0 & \times & 0
\end{bmatrix},
\:\:\:\:\:
Y_2 =
\begin{bmatrix}
0 & \times & 0 \\
\times & 0 & 0 \\
0 & 0 & 0
\end{bmatrix},
\:\:\:\:\:
X_1 =
\begin{bmatrix}
0 & 0 & 0 \\
0 & \times & \times \\
\times & 0 & 0
\end{bmatrix},
\:\:\:\:\:
X_2 =
\begin{bmatrix}
0 & \times & \times \\
\times & 0 & 0 \\
0 & 0 & 0
\end{bmatrix},
\:\:\:\:\:
N =
\begin{bmatrix}
0 & 0 & 0 \\
0 & \times & \times \\
0 & \times & \times
\end{bmatrix},
\end{equation}
\begin{equation}
    W_{10} = \mathrm{diag}(1,\e^{4\i\pi/3},\e^{2\i\pi/3}), \:\:\:\:\:\: W_{5} = \mathrm{diag}(\e^{4\i\pi/3},\e^{2\i\pi/3},\e^{2\i\pi/3}),
\end{equation}
\begin{equation}
    \e^{\i\varphi_{5}} = 1, \:\:\:\:\:\: \e^{\i\varphi_{45}} = \e^{2\i\pi/3}, \:\:\:\:\:\: \e^{\i\varphi_{15}} = \e^{2\i\pi/3}.
\end{equation}

\noindent
Case $D_2$ ($\Z_3$ symmetry):
\begin{equation}
Y_1 =
\begin{bmatrix}
\times & 0 & 0 \\
0 & 0 & \times \\
0 & \times & 0
\end{bmatrix},
\:\:\:\:\:
Y_2 =
\begin{bmatrix}
0 & \times & 0 \\
\times & 0 & 0 \\
0 & 0 & 0
\end{bmatrix},
\:\:\:\:\:
X_1 =
\begin{bmatrix}
\times & 0 & 0 \\
0 & 0 & 0 \\
0 & \times & \times
\end{bmatrix},
\:\:\:\:\:
X_2 =
\begin{bmatrix}
0 & 0 & 0 \\
0 & \times & \times \\
\times & 0 & 0
\end{bmatrix},
\:\:\:\:\:
N =
\begin{bmatrix}
0 & 0 & 0 \\
0 & \times & \times \\
0 & \times & \times
\end{bmatrix},
\end{equation}
\begin{equation}
    W_{10} = \mathrm{diag}(\e^{2\i\pi/3},\e^{4\i\pi/3},1), \:\:\:\:\:\: W_{5} = \mathrm{diag}(1,\e^{2\i\pi/3},\e^{2\i\pi/3}),
\end{equation}
\begin{equation}
    \e^{\i\varphi_{5}} = \e^{2\i\pi/3}, \:\:\:\:\:\: \e^{\i\varphi_{45}} = 1, \:\:\:\:\:\: \e^{\i\varphi_{15}} = \e^{2\i\pi/3}.
\end{equation}

\noindent
Case $D_3$ ($\Z_3$ symmetry):
\begin{equation}
Y_1 =
\begin{bmatrix}
\times & 0 & 0 \\
0 & 0 & \times \\
0 & \times & 0
\end{bmatrix},
\:\:\:\:\:
Y_2 =
\begin{bmatrix}
0 & \times & 0 \\
\times & 0 & 0 \\
0 & 0 & 0
\end{bmatrix},
\:\:\:\:\:
X_1 =
\begin{bmatrix}
0 & 0 & 0 \\
\times & 0 & 0 \\
0 & \times & \times
\end{bmatrix},
\:\:\:\:\:
X_2 =
\begin{bmatrix}
\times & 0 & 0 \\
0 & \times & \times \\
0 & 0 & 0
\end{bmatrix},
\:\:\:\:\:
N =
\begin{bmatrix}
0 & 0 & 0 \\
0 & \times & \times \\
0 & \times & \times
\end{bmatrix},
\end{equation}
\begin{equation}
    W_{10} = \mathrm{diag}(\e^{2\i\pi/3},\e^{4\i\pi/3},1), \:\:\:\:\:\: W_{5} = \mathrm{diag}(\e^{4\i\pi/3},\e^{2\i\pi/3},\e^{2\i\pi/3}),
\end{equation}
\begin{equation}
    \e^{\i\varphi_{5}} = \e^{2\i\pi/3}, \:\:\:\:\:\: \e^{\i\varphi_{45}} = 1, \:\:\:\:\:\: \e^{\i\varphi_{15}} = \e^{2\i\pi/3}.
\end{equation}

\noindent
Case $D_4$ ($\Z_3$ symmetry):
\begin{equation}
Y_1 =
\begin{bmatrix}
\times & 0 & 0 \\
0 & 0 & \times \\
0 & \times & 0
\end{bmatrix},
\:\:\:\:\:
Y_2 =
\begin{bmatrix}
0 & \times & 0 \\
\times & 0 & 0 \\
0 & 0 & 0
\end{bmatrix},
\:\:\:\:\:
X_1 =
\begin{bmatrix}
0 & \times & \times \\
0 & 0 & 0 \\
\times & 0 & 0
\end{bmatrix},
\:\:\:\:\:
X_2 =
\begin{bmatrix}
0 & 0 & 0 \\
\times & 0 & 0 \\
0 & \times & \times
\end{bmatrix},
\:\:\:\:\:
N =
\begin{bmatrix}
0 & 0 & 0 \\
0 & \times & \times \\
0 & \times & \times
\end{bmatrix},
\end{equation}
\begin{equation}
    W_{10} = \mathrm{diag}(\e^{2\i\pi/3},\e^{4\i\pi/3},1), \:\:\:\:\:\: W_{5} = \mathrm{diag}(\e^{2\i\pi/3},1,1),
\end{equation}
\begin{equation}
    \e^{\i\varphi_{5}} = \e^{2\i\pi/3}, \:\:\:\:\:\: \e^{\i\varphi_{45}} = 1, \:\:\:\:\:\: \e^{\i\varphi_{15}} = 1.
\end{equation}

\noindent
Case $C_1^{\mathrm{I}}$ ($\Z_4$ symmetry):
\begin{equation}
Y_1 =
\begin{bmatrix}
0 & 0 & 0 \\
0 & \times & 0 \\
0 & 0 & \times
\end{bmatrix},
\:\:\:\:\:
Y_2 =
\begin{bmatrix}
0 & 0 & \times \\
0 & 0 & 0 \\
\times & 0 & 0
\end{bmatrix},
\:\:\:\:\:
X_1 =
\begin{bmatrix}
0 & \times & \times \\
\times & 0 & 0 \\
0 & 0 & 0
\end{bmatrix},
\:\:\:\:\:
X_2 =
\begin{bmatrix}
\times & 0 & 0 \\
0 & 0 & 0 \\
0 & \times & \times
\end{bmatrix},
\:\:\:\:\:
N =
\begin{bmatrix}
0 & 0 & 0 \\
0 & \times & \times \\
0 & \times & \times
\end{bmatrix},
\end{equation}
\begin{equation}
    W_{10} = \mathrm{diag}(\i, 1, -1), \:\:\:\:\:\: W_{5} = \mathrm{diag}(1, -\i, -\i),
\end{equation}
\begin{equation}
    \e^{\i\varphi_{5}} = 1, \:\:\:\:\:\: \e^{\i\varphi_{45}} = \i, \:\:\:\:\:\: \e^{\i\varphi_{15}} = -1.
\end{equation}

\noindent
Case $C_1^{\mathrm{II}}$ ($\Z_4$ symmetry):
\begin{equation}
Y_1 =
\begin{bmatrix}
0 & 0 & 0 \\
0 & \times & 0 \\
0 & 0 & \times
\end{bmatrix},
\:\:\:\:\:
Y_2 =
\begin{bmatrix}
0 & 0 & \times \\
0 & 0 & 0 \\
\times & 0 & 0
\end{bmatrix},
\:\:\:\:\:
X_1 =
\begin{bmatrix}
0 & 0 & \times \\
\times & 0 & 0 \\
0 & 0 & 0
\end{bmatrix},
\:\:\:\:\:
X_2 =
\begin{bmatrix}
\times & 0 & 0 \\
0 & \times & 0 \\
0 & 0 & \times
\end{bmatrix},
\:\:\:\:\:
N =
\begin{bmatrix}
0 & 0 & 0 \\
0 & \times & 0 \\
0 & 0 & \times
\end{bmatrix},
\end{equation}
\begin{equation}
    W_{10} = \mathrm{diag}(\i, 1, -1), \:\:\:\:\:\: W_{5} = \mathrm{diag}(1, \i, -\i),
\end{equation}
\begin{equation}
    \e^{\i\varphi_{5}} = 1, \:\:\:\:\:\: \e^{\i\varphi_{45}} = \i, \:\:\:\:\:\: \e^{\i\varphi_{15}} = -1.
\end{equation}

\noindent
Case $C_1^{\mathrm{III}}$ ($\Z_4$ symmetry):
\begin{equation}
Y_1 =
\begin{bmatrix}
\times & 0 & 0 \\
0 & 0 & \times \\
0 & \times & 0
\end{bmatrix},
\:\:\:\:\:
Y_2 =
\begin{bmatrix}
0 & \times & 0 \\
\times & 0 & 0 \\
0 & 0 & 0
\end{bmatrix},
\:\:\:\:\:
X_1 =
\begin{bmatrix}
0 & 0 & \times \\
\times & 0 & 0 \\
0 & 0 & 0
\end{bmatrix},
\:\:\:\:\:
X_2 =
\begin{bmatrix}
\times & 0 & 0 \\
0 & \times & 0 \\
0 & 0 & \times
\end{bmatrix},
\:\:\:\:\:
N =
\begin{bmatrix}
0 & 0 & 0 \\
0 & \times & 0 \\
0 & 0 & \times
\end{bmatrix},
\end{equation}
\begin{equation}
    W_{10} = \mathrm{diag}(\i, -1, 1), \:\:\:\:\:\: W_{5} = \mathrm{diag}(1, -\i, \i),
\end{equation}
\begin{equation}
    \e^{\i\varphi_{5}} = -1, \:\:\:\:\:\: \e^{\i\varphi_{45}} = \i, \:\:\:\:\:\: \e^{\i\varphi_{15}} = -1.
\end{equation}

\noindent
Case $C_2^{\mathrm{I}}$ ($\Z_4$ symmetry):
\begin{equation}
Y_1 =
\begin{bmatrix}
0 & 0 & 0 \\
0 & \times & \times \\
0 & \times & \times
\end{bmatrix},
\:\:\:\:\:
Y_2 =
\begin{bmatrix}
0 & \times & \times \\
\times & 0 & 0 \\
\times & 0 & 0
\end{bmatrix},
\:\:\:\:\:
X_1 =
\begin{bmatrix}
\times & 0 & 0 \\
0 & \times & 0 \\
0 & \times & 0
\end{bmatrix},
\:\:\:\:\:
X_2 =
\begin{bmatrix}
0 & 0 & \times \\
\times & 0 & 0 \\
\times & 0 & 0
\end{bmatrix},
\:\:\:\:\:
N =
\begin{bmatrix}
0 & 0 & 0 \\
0 & \times & 0 \\
0 & 0 & \times
\end{bmatrix},
\end{equation}
\begin{equation}
    W_{10} = \mathrm{diag}(\i, 1, 1), \:\:\:\:\:\: W_{5} = \mathrm{diag}(-\i, 1, -1),
\end{equation}
\begin{equation}
    \e^{\i\varphi_{5}} = 1, \:\:\:\:\:\: \e^{\i\varphi_{45}} = -\i, \:\:\:\:\:\: \e^{\i\varphi_{15}} = 1.
\end{equation}

\noindent
Case $C_2^{\mathrm{II}}$ ($\Z_4$ symmetry):
\begin{equation}
Y_1 =
\begin{bmatrix}
0 & 0 & 0 \\
0 & \times & 0 \\
0 & 0 & \times
\end{bmatrix},
\:\:\:\:\:
Y_2 =
\begin{bmatrix}
0 & 0 & \times \\
0 & 0 & 0 \\
\times & 0 & 0
\end{bmatrix},
\:\:\:\:\:
X_1 =
\begin{bmatrix}
\times & 0 & 0 \\
0 & \times & \times \\
0 & 0 & 0
\end{bmatrix},
\:\:\:\:\:
X_2 =
\begin{bmatrix}
0 & \times & \times \\
0 & 0 & 0 \\
\times & 0 & 0
\end{bmatrix},
\:\:\:\:\:
N =
\begin{bmatrix}
0 & 0 & 0 \\
0 & \times & \times \\
0 & \times & \times
\end{bmatrix},
\end{equation}
\begin{equation}
    W_{10} = \mathrm{diag}(\i, 1, -1), \:\:\:\:\:\: W_{5} = \mathrm{diag}(-\i, 1, 1),
\end{equation}
\begin{equation}
    \e^{\i\varphi_{5}} = 1, \:\:\:\:\:\: \e^{\i\varphi_{45}} = \i, \:\:\:\:\:\: \e^{\i\varphi_{15}} = 1.
\end{equation}

\noindent
Case $C_2^{\mathrm{III}}$ ($\Z_4$ symmetry):
\begin{equation}
Y_1 =
\begin{bmatrix}
0 & 0 & 0 \\
0 & \times & 0 \\
0 & 0 & \times
\end{bmatrix},
\:\:\:\:\:
Y_2 =
\begin{bmatrix}
0 & 0 & \times \\
0 & 0 & 0 \\
\times & 0 & 0
\end{bmatrix},
\:\:\:\:\:
X_1 =
\begin{bmatrix}
\times & 0 & 0 \\
0 & \times & 0 \\
0 & 0 & \times
\end{bmatrix},
\:\:\:\:\:
X_2 =
\begin{bmatrix}
0 & \times & 0 \\
0 & 0 & 0 \\
\times & 0 & 0
\end{bmatrix},
\:\:\:\:\:
N =
\begin{bmatrix}
0 & 0 & 0 \\
0 & \times & 0 \\
0 & 0 & \times
\end{bmatrix},
\end{equation}
\begin{equation}
    W_{10} = \mathrm{diag}(-\i, -1, 1), \:\:\:\:\:\: W_{5} = \mathrm{diag}(\i, -1, 1),
\end{equation}
\begin{equation}
    \e^{\i\varphi_{5}} = 1, \:\:\:\:\:\: \e^{\i\varphi_{45}} = \i, \:\:\:\:\:\: \e^{\i\varphi_{15}} = 1.
\end{equation}

\noindent
Case $C_2^{\mathrm{IV}}$ ($\Z_4$ symmetry):
\begin{equation}
Y_1 =
\begin{bmatrix}
\times & 0 & 0 \\
0 & 0 & \times \\
0 & \times & 0
\end{bmatrix},
\:\:\:\:\:
Y_2 =
\begin{bmatrix}
0 & \times & 0 \\
\times & 0 & 0 \\
0 & 0 & 0
\end{bmatrix},
\:\:\:\:\:
X_1 =
\begin{bmatrix}
\times & 0 & 0 \\
0 & \times & 0 \\
0 & 0 & \times
\end{bmatrix},
\:\:\:\:\:
X_2 =
\begin{bmatrix}
0 & \times & 0 \\
0 & 0 & 0 \\
\times & 0 & 0
\end{bmatrix},
\:\:\:\:\:
N =
\begin{bmatrix}
0 & 0 & 0 \\
0 & \times & 0 \\
0 & 0 & \times
\end{bmatrix},
\end{equation}
\begin{equation}
    W_{10} = \mathrm{diag}(\i, -1, 1), \:\:\:\:\:\: W_{5} = \mathrm{diag}(\i, 1, -1),
\end{equation}
\begin{equation}
    \e^{\i\varphi_{5}} = -1, \:\:\:\:\:\: \e^{\i\varphi_{45}} = \i, \:\:\:\:\:\: \e^{\i\varphi_{15}} = 1.
\end{equation}

\noindent
Case $A_1'$ ($\U(1)$ symmetry):
\begin{equation}
Y_1 =
\begin{bmatrix}
\times & \times & \times \\
\times & \times & \times \\
\times & \times & \times
\end{bmatrix},
\:\:\:\:\:
Y_2 =
\begin{bmatrix}
0 & 0 & 0 \\
0 & 0 & 0 \\
0 & 0 & 0
\end{bmatrix},
\:\:\:\:\:
X_1 =
\begin{bmatrix}
0 & \times & \times \\
0 & \times & \times \\
0 & \times & \times
\end{bmatrix},
\:\:\:\:\:
X_2 =
\begin{bmatrix}
\times & 0 & 0 \\
\times & 0 & 0 \\
\times & 0 & 0
\end{bmatrix},
\:\:\:\:\:
N =
\begin{bmatrix}
0 & 0 & 0 \\
0 & \times & \times \\
0 & \times & \times
\end{bmatrix},
\end{equation}
\begin{equation}
    W_{10} = \mathrm{diag}(1, 1, 1), \:\:\:\:\:\: W_{5} = \mathrm{diag}(\e^{\i\phi}, 1, 1),
\end{equation}
\begin{equation}
    \e^{\i\varphi_{5}} = 1, \:\:\:\:\:\: \e^{\i\varphi_{45}} = \e^{\i\phi}, \:\:\:\:\:\: \e^{\i\varphi_{15}} = 1.
\end{equation}

\noindent
Case $A_2'$ ($\U(1)$ symmetry):
\begin{equation}
\label{eq:A2pmats}
Y_1 =
\begin{bmatrix}
\times & \times & \times \\
\times & \times & \times \\
\times & \times & \times
\end{bmatrix},
\:\:\:\:\:
Y_2 =
\begin{bmatrix}
0 & 0 & 0 \\
0 & 0 & 0 \\
0 & 0 & 0
\end{bmatrix},
\:\:\:\:\:
X_1 =
\begin{bmatrix}
\times & 0 & 0 \\
\times & 0 & 0 \\
\times & 0 & 0
\end{bmatrix},
\:\:\:\:\:
X_2 =
\begin{bmatrix}
0 & \times & \times \\
0 & \times & \times \\
0 & \times & \times
\end{bmatrix},
\:\:\:\:\:
N =
\begin{bmatrix}
0 & 0 & 0 \\
0 & \times & \times \\
0 & \times & \times
\end{bmatrix},
\end{equation}
\begin{equation}
    W_{10} = \mathrm{diag}(1, 1, 1), \:\:\:\:\:\: W_{5} = \mathrm{diag}(1, \e^{\i\phi}, \e^{\i\phi}),
\end{equation}
\begin{equation}
\label{eq:A2pscals}
    \e^{\i\varphi_{5}} = 1, \:\:\:\:\:\: \e^{\i\varphi_{45}} = \e^{\i\phi}, \:\:\:\:\:\: \e^{\i\varphi_{15}} = \e^{-2\i\phi}.
\end{equation}

\noindent
Case $B_2'$ ($\U(1)$ symmetry):
\begin{equation}
Y_1 =
\begin{bmatrix}
0 & 0 & 0 \\
0 & \times & \times \\
0 & \times & \times
\end{bmatrix},
\:\:\:\:\:
Y_2 =
\begin{bmatrix}
0 & \times & \times \\
\times & 0 & 0 \\
\times & 0 & 0
\end{bmatrix},
\:\:\:\:\:
X_1 =
\begin{bmatrix}
\times & \times & \times \\
0 & 0 & 0 \\
0 & 0 & 0
\end{bmatrix},
\:\:\:\:\:
X_2 =
\begin{bmatrix}
0 & 0 & 0 \\
\times & \times & \times \\
\times & \times & \times
\end{bmatrix},
\:\:\:\:\:
N =
\begin{bmatrix}
\times & \times & \times \\
\times & \times & \times \\
\times & \times & \times
\end{bmatrix},
\end{equation}
\begin{equation}
    W_{10} = \mathrm{diag}(\e^{\i\phi}, 1, 1), \:\:\:\:\:\: W_{5} = \mathrm{diag}(\e^{-\i\phi}, \e^{-\i\phi}, \e^{-\i\phi}),
\end{equation}
\begin{equation}
    \e^{\i\varphi_{5}} = 1, \:\:\:\:\:\: \e^{\i\varphi_{45}} = \e^{-\i\phi}, \:\:\:\:\:\: \e^{\i\varphi_{15}} = \e^{2\i\phi}.
\end{equation}

\noindent
Case $C_1'$ ($\U(1)$ symmetry):
\begin{equation}
Y_1 =
\begin{bmatrix}
\times & 0 & 0 \\
0 & 0 & \times \\
0 & \times & 0
\end{bmatrix},
\:\:\:\:\:
Y_2 =
\begin{bmatrix}
0 & \times & 0 \\
\times & 0 & 0 \\
0 & 0 & 0
\end{bmatrix},
\:\:\:\:\:
X_1 =
\begin{bmatrix}
0 & \times & \times \\
\times & 0 & 0 \\
0 & 0 & 0
\end{bmatrix},
\:\:\:\:\:
X_2 =
\begin{bmatrix}
\times & 0 & 0 \\
0 & 0 & 0 \\
0 & \times & \times
\end{bmatrix},
\:\:\:\:\:
N =
\begin{bmatrix}
0 & 0 & 0 \\
0 & \times & \times \\
0 & \times & \times
\end{bmatrix},
\end{equation}
\begin{equation}
    W_{10} = \mathrm{diag}(1, \e^{-\i\phi}, \e^{\i\phi}), \:\:\:\:\:\: W_{5} = \mathrm{diag}(\e^{\i\phi}, 1, 1),
\end{equation}
\begin{equation}
    \e^{\i\varphi_{5}} = 1, \:\:\:\:\:\: \e^{\i\varphi_{45}} = \e^{\i\phi}, \:\:\:\:\:\: \e^{\i\varphi_{15}} = 1.
\end{equation}

\noindent
Case $C_1''$ ($\U(1)$ symmetry):
\begin{equation}
Y_1 =
\begin{bmatrix}
0 & 0 & 0 \\
0 & \times & \times \\
0 & \times & \times
\end{bmatrix},
\:\:\:\:\:
Y_2 =
\begin{bmatrix}
0 & \times & \times \\
\times & 0 & 0 \\
\times & 0 & 0
\end{bmatrix},
\:\:\:\:\:
X_1 =
\begin{bmatrix}
0 & \times & \times \\
0 & 0 & 0 \\
0 & 0 & 0
\end{bmatrix},
\:\:\:\:\:
X_2 =
\begin{bmatrix}
\times & 0 & 0 \\
0 & \times & \times \\
0 & \times & \times
\end{bmatrix},
\:\:\:\:\:
N =
\begin{bmatrix}
0 & 0 & 0 \\
0 & \times & \times \\
0 & \times & \times
\end{bmatrix},
\end{equation}
\begin{equation}
    W_{10} = \mathrm{diag}(\e^{\i\phi}, 1, 1), \:\:\:\:\:\: W_{5} = \mathrm{diag}(\e^{-2\i\phi}, \e^{-\i\phi}, \e^{-\i\phi}),
\end{equation}
\begin{equation}
    \e^{\i\varphi_{5}} = 1, \:\:\:\:\:\: \e^{\i\varphi_{45}} = \e^{-\i\phi}, \:\:\:\:\:\: \e^{\i\varphi_{15}} = \e^{2\i\phi}.
\end{equation}

\noindent
Case $C_1'''$ ($\U(1)$ symmetry):
\begin{equation}
Y_1 =
\begin{bmatrix}
0 & 0 & 0 \\
0 & \times & \times \\
0 & \times & \times
\end{bmatrix},
\:\:\:\:\:
Y_2 =
\begin{bmatrix}
0 & \times & \times \\
\times & 0 & 0 \\
\times & 0 & 0
\end{bmatrix},
\:\:\:\:\:
X_1 =
\begin{bmatrix}
0 & \times & \times \\
\times & 0 & 0 \\
\times & 0 & 0
\end{bmatrix},
\:\:\:\:\:
X_2 =
\begin{bmatrix}
0 & 0 & 0 \\
0 & \times & \times \\
0 & \times & \times
\end{bmatrix},
\:\:\:\:\:
N =
\begin{bmatrix}
0 & 0 & 0 \\
0 & \times & \times \\
0 & \times & \times
\end{bmatrix},
\end{equation}
\begin{equation}
    W_{10} = \mathrm{diag}(\e^{\i\phi}, 1, 1), \:\:\:\:\:\: W_{5} = \mathrm{diag}(1, \e^{-\i\phi}, \e^{-\i\phi}),
\end{equation}
\begin{equation}
    \e^{\i\varphi_{5}} = 1, \:\:\:\:\:\: \e^{\i\varphi_{45}} = \e^{-\i\phi}, \:\:\:\:\:\: \e^{\i\varphi_{15}} = \e^{2\i\phi}.
\end{equation}

\noindent
Case $C_2'$ ($\U(1)$ symmetry):
\begin{equation}
Y_1 =
\begin{bmatrix}
\times & 0 & 0 \\
0 & 0 & \times \\
0 & \times & 0
\end{bmatrix},
\:\:\:\:\:
Y_2 =
\begin{bmatrix}
0 & \times & 0 \\
\times & 0 & 0 \\
0 & 0 & 0
\end{bmatrix},
\:\:\:\:\:
X_1 =
\begin{bmatrix}
\times & 0 & 0 \\
0 & \times & \times \\
0 & 0 & 0
\end{bmatrix},
\:\:\:\:\:
X_2 =
\begin{bmatrix}
0 & \times & \times \\
0 & 0 & 0 \\
\times & 0 & 0
\end{bmatrix},
\:\:\:\:\:
N =
\begin{bmatrix}
0 & 0 & 0 \\
0 & \times & \times \\
0 & \times & \times
\end{bmatrix},
\end{equation}
\begin{equation}
    W_{10} = \mathrm{diag}(1, \e^{-\i\phi}, \e^{\i\phi}), \:\:\:\:\:\: W_{5} = \mathrm{diag}(1, \e^{\i\phi}, \e^{\i\phi}),
\end{equation}
\begin{equation}
    \e^{\i\varphi_{5}} = 1, \:\:\:\:\:\: \e^{\i\varphi_{45}} = \e^{\i\phi}, \:\:\:\:\:\: \e^{\i\varphi_{15}} = \e^{-2\i\phi}.
\end{equation}

\noindent
Case $C_2''$ ($\U(1)$ symmetry):
\begin{equation}
Y_1 =
\begin{bmatrix}
0 & 0 & 0 \\
0 & \times & \times \\
0 & \times & \times
\end{bmatrix},
\:\:\:\:\:
Y_2 =
\begin{bmatrix}
0 & \times & \times \\
\times & 0 & 0 \\
\times & 0 & 0
\end{bmatrix},
\:\:\:\:\:
X_1 =
\begin{bmatrix}
\times & 0 & 0 \\
0 & \times & \times \\
0 & \times & \times
\end{bmatrix},
\:\:\:\:\:
X_2 =
\begin{bmatrix}
0 & 0 & 0 \\
\times & 0 & 0 \\
\times & 0 & 0
\end{bmatrix},
\:\:\:\:\:
N =
\begin{bmatrix}
0 & 0 & 0 \\
0 & \times & \times \\
0 & \times & \times
\end{bmatrix},
\end{equation}
\begin{equation}
    W_{10} = \mathrm{diag}(\e^{\i\phi}, 1, 1), \:\:\:\:\:\: W_{5} = \mathrm{diag}(\e^{-\i\phi}, 1, 1),
\end{equation}
\begin{equation}
    \e^{\i\varphi_{5}} = 1, \:\:\:\:\:\: \e^{\i\varphi_{45}} = \e^{-\i\phi}, \:\:\:\:\:\: \e^{\i\varphi_{15}} = 1.
\end{equation}

\section{Non-Abelian symmetry cases}
\label{app:NA}

Given the experimentally motivated requirements, all possible Abelian symmetry cases have been determined (see section~\ref{ch:cases}). In section~\ref{ch:flavor_symmetry}, it was assumed that all transformation matrices $W$ belonging to a flavor symmetry are simultaneously diagonalizable, which rules out non-Abelian symmetries. This must be justified, since such symmetries are, in principle, an equally valid option. In this appendix, we show that no non-Abelian symmetries exist that satisfy our requirements. The argument relies on an attempt to explicitly construct the transformation matrices, which invariably leads to a violation of at least one of our requirements.

Every non-Abelian symmetry will have multiple transformations (that is, sets of $W_5$ and $W_{10}$ together with phases) that do not all commute. As $W_5$ and $W_{10}$ are unitary and act in unrelated bases, any set of them may be diagonalized with the proper choice of 10-plet and 5-plet generation bases. Since the sets do not all commute, all sets may not be diagonalized at the same time. Therefore, every non-Abelian symmetry case is a subcase of an Abelian case with diagonal $W_5$ and $W_{10}$, where at least one additional transformation set $W_{10}'$ and $W_5'$ that does not commute with the diagonal set has been introduced. All possible symmetry cases with diagonal $W$ matrices have already been found, and thus, any non-Abelian cases must be subcases of them.

In all found cases, a 5-plet generation basis may be chosen such that the matrix $N$ is diagonal, while keeping $W_5$ diagonal and not affecting the shape of the other coupling matrices. In accordance with our assumptions, the three elements in $N$ must be different from each other, since they correspond to the neutrino masses. The invariance of $N$ under the additional transformation $W_5'$ is enforced by eq.~\eqref{invariantN}, which may be rephrased as
\begin{equation}
    \e^{\i\varphi_{15}'}NW_5' = W_5'^* N.
\end{equation}
Evaluating this condition element-wisely is straightforward, since $N$ is diagonal in the chosen basis. Taking the absolute value of each element yields
\begin{equation}
    |N^{(ii)}W_5'^{(ij)}| = |W_5'^{(ij)}N^{(jj)}|
    \label{NWWN}
\end{equation}
for all $i,j = 1,2,3$. If $i=j$, eq.~\eqref{NWWN} is trivially true. However, if $i\neq j$, then the only two options are
\begin{equation}
    N^{(ii)} = N^{(jj)} \:\:\:\:\: \mathrm{or} \:\:\:\:\: W_5'^{(ij)} = 0.
\end{equation}
Since the diagonal elements of $N$ are all different, only the second option is possible. Therefore, $W_5'$ must be diagonal and commutes with the original diagonal $W_5$. Now, it must be checked whether $W_{10}'$ will commute with the original $W_{10}$ or not.

For a number of cases, $W_{10}$ is the identity matrix and will always commute with a new $W_{10}'$. It is therefore impossible to introduce a non-Abelian subcase to these cases. In all other cases, a generation basis for the 10-plet may be chosen such that
\begin{equation}
    Y_2 =
    \begin{bmatrix}
    0 & 0 & b \\
    0 & 0 & 0 \\
    -b & 0 & 0
    \end{bmatrix},
    \label{Y2_option}
\end{equation}
while keeping $W_{10}$ diagonal and not affecting the shape of the other coupling matrices. If $b=0$, then there are two different possibilities, depending on the specific case. In some cases, it is impossible to fit three different up-type quark masses, which means that $b \neq 0$. In the remaining cases, if $b=0$, then another basis may be chosen such that $Y_1$ is diagonal with three different elements, corresponding to the three up-type quark masses. The same arguments as for $N$ may then be repeated, meaning that $W_{10}'$ commutes with $W_{10}$, and the new symmetry is Abelian. Since a non-Abelian symmetry is sought, this again requires that $b \neq 0$.

To better see the allowed values of $W_{10}'$, eq.~\eqref{invariantY2} is rephrased as
\begin{equation}
    \e^{\i\varphi_{45}'} Y_2W_{10}' = W_{10}'^* Y_2.
    \label{YWWY}
\end{equation}
Using eq.~\eqref{Y2_option} and carrying out the matrix multiplication, eq.~\eqref{YWWY} becomes
\begin{equation}
    b \e^{\i\varphi_{45}'}
    \begin{bmatrix}
        w_{31} & w_{32} & w_{33} \\
        0 & 0 & 0 \\
        -w_{11} & -w_{12} -& w_{13}
    \end{bmatrix}
    = b
    \begin{bmatrix}
        -w_{13}^* & 0 & w_{11}^* \\
        -w_{23}^* & 0 & w_{21}^* \\
        -w_{33}^* & 0 & w_{31}^*
    \end{bmatrix},
\end{equation}
where $w_{ij}$ are the elements of $W_{10}'$. Any allowed $W_{10}'$ must be in the form
\begin{equation}
    W_{10}' =
    \begin{bmatrix}
    w_{11} & 0 & w_{13} \\
    0 & w_{22} & 0 \\
    -\e^{-\i\varphi_{45}'}w_{13}^* & 0 & \e^{-\i\varphi_{45}'}w_{11}^*
    \end{bmatrix},
\end{equation}
where $w_{13} \neq 0$ if the symmetry is to be non-Abelian. However, by requiring $X_{1,2}$ to be invariant under such a transformation, in all found cases the down-type quark and charged lepton mass matrices must be singular. Thus, any non-Abelian subcase of the found cases will violate the chosen requirements.

The arguments above show that given the requirements used and the Abelian cases they allow, it is impossible to introduce a non-Abelian case.

\bibliographystyle{apsrev4-1}
\bibliography{ref.bib}

\end{document}